\pdfoutput=1

\documentclass{PoS}
\usepackage{amsmath}
\usepackage{wrapfig}

\DeclareRobustCommand*\diff[2][]{%
   \mathop{\mathrm{d}^{#1} \mskip-0.2\thinmuskip #2}
   \nolimits
}
\newcommand{\VC}[1]{%
  \begin{tabular}[c]{l}%
    #1%
  \end{tabular}
}
\newcommand{\T}[1]{\boldsymbol{#1}_{\text{T}}}
\title{Do fragmentation functions in factorization theorems correctly
  treat non-perturbative effects?
}

\ShortTitle{Fragmentation and non-perturbative effects}

\author{\speaker{John Collins}\\
        104 Davey Lab, Penn State University, University Park PA
            16802, USA\\
        E-mail: \email{jcc8@psu.edu}}

\abstract{Current all-orders proofs of factorization of hard processes
  are made by extracting the leading power behavior of Feynman graphs,
  i.e., by extracting asymptotics strictly order-by-order in
  perturbation theory.  The resulting parton densities and
  fragmentation functions include non-perturbative effects.  I show
  how there are missing elements in the proofs; these are related to
  and exemplified by string and cluster models of hadronization.  The
  proofs rely on large rapidity differences between different parts of
  graphs for the process; but in reality large rapidity gaps are
  filled in.}

\FullConference{QCD Evolution 2016\\
		May 30-June 03, 2016\\
		National Institute for Subatomic Physics (Nikhef), Amsterdam} 

\begin{document}

\section{Introduction}

Factorization theorems (of both the collinear and the
transverse-momentum-dependent (TMD) kinds) are intended to be
applicable in full QCD. They apply to a number of important reactions
to within an error suppressed by a power of a large scale.  As well as
perturbatively calculable components (notably hard-scattering factors,
evolution kernels), the factorization properties include factors like
parton densities and fragmentation functions that are intended to
incorporate the relevant non-perturbative information.

However, the actual published proofs are normally made by a systematic
analysis of the leading power in each order of perturbation theory.
Therefore, since we claim that factorization properties do incorporate
genuine first-principles predictions of QCD, we must ask how well we
know that factorization is justified in full QCD, and not merely in
order-by-order perturbation theory.  This issue has been recognized
for a long time, e.g., \cite{Webber:1999ui}.

As we will see, there are many elements in an order-by-order
derivation of factorization that do appear to have a validity beyond
pure perturbation theory.  The same applies to the parton-model
intuition that motivated the factorization approach.

In this paper, I will show explicitly that, in contrast, there are
certain parts of the derivation that clash with what we know of
hadronization in the final state in QCD in hard collisions.  These are
the phenomena that are approximated in Monte-Carlo event generators
under the heading of string or cluster hadronization
\cite{Webber:1999ui}.  The primary issue is that existing perturbative
proofs of factorization need large rapidity differences between the
different momentum classes considered, but hadronization fills in the
rapidity gaps, as is known experimentally.  The filling in of rapidity
gaps can even be modeled within perturbation theory when the order of
important Feynman graphs increases logarithmically with energy.

These results indicate that some physically important phenomena are
not properly treated within the normal perturbatively-based
derivations of factorization.  This does not necessarily imply that
factorization actually fails.  But it does imply that our current
understanding is incomplete.

\section{Factorization and the importance of non-perturbative
  components}

\subsection{Basics of factorization}

Hard-scattering factorization applies to many kinds of cross section
with a large scale $Q$ (e.g., deeply inelastic scattering (DIS)), and
has a typical form of a hard scattering convoluted with parton
densities (pdfs) and/or fragmentation functions (ffs):
\begin{equation}
    \sigma = \mbox{hard scattering} \otimes \mbox{pdfs and/or ffs}
             ~+~ \mbox{power-suppressed}.
\end{equation}
Pdfs and ffs have scale dependence that is governed by evolution
equations, e.g., that of DGLAP.
The predictive power given to QCD by factorization properties arises
from the following:
\begin{itemize}

\item The hard scattering, the DGLAP kernels, etc can be usefully
  calculated by low-order perturbative calculations in powers of a
  small effective coupling $\alpha_s$.

\item In contrast, pdfs and ffs cannot be computed perturbatively.
  But they can be measured from a limited set of data over a limited
  range of $Q$, as can the scale parameter $\Lambda_{\rm QCD}$ for the
  evolution of $\alpha_s$.

\item Then, after evolution is allowed for, the universality of pdfs,
  ffs, etc enables factorization to give predictions for many other
  processes at all high enough $Q$.  Similarly it also gives
  predictions for cross sections for the processes used to measure the
  pdfs etc, but at different $Q$ values than those used for their
  measurement.

\end{itemize}

\subsection{Non-perturbative reasoning in coordinate-space motivates
  factorization/parton ideas}

\begin{figure}
\setlength\tabcolsep{0.015\textwidth}
\centering
\begin{tabular}{cc}
      \VC{\includegraphics[scale=0.4]{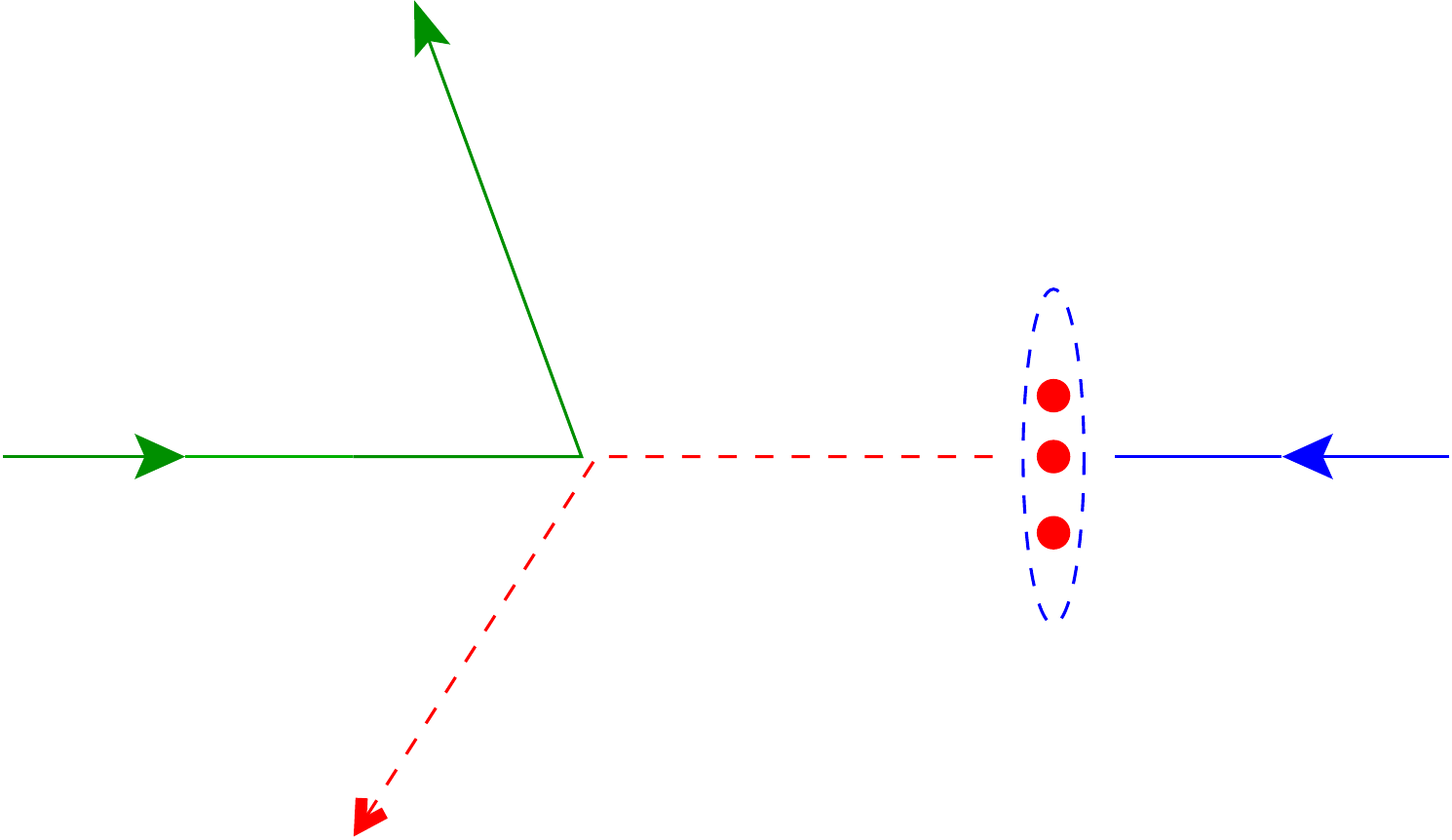}}
   &
      \VC{\includegraphics[scale=0.9]{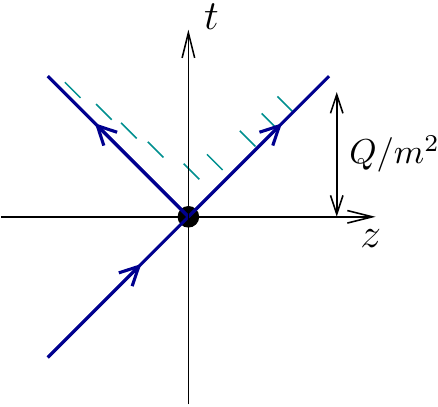}}
\\
   \begin{minipage}[t]{0.45\textwidth}
     \caption{DIS in parton model, with a proton coming from the right
       and an electron from the left.  The hard scattering is off one
       constituent (red, dashed) of the proton.}
      \label{fig:DIS.pic}
   \end{minipage}
   &
   \begin{minipage}[t]{0.45\textwidth}
      \centering
      \caption{Space-time view corresponding to
        Fig.\ \protect\ref{fig:DIS.pic} in the Breit frame.}
      \label{fig:DIS.sp.time}
   \end{minipage}
\end{tabular}
\end{figure}

The motivation that led to factorization theorems in QCD was the
parton model.  In the case of DIS, $e+p \to e' +X$, the parton model
assumptions are illustrated in Fig.\ \ref{fig:DIS.pic}, in the overall
center-of-mass frame.  There is a fast-moving incoming
Lorentz-contracted, time-dilated proton (or other hadron).  The
incoming electron scatters through a wide-angle hard scattering from a
single constituent of the hadron, over a small distance scale of order
$1/Q$.  This is much smaller than the time scale $Q/m^2$ for valence
processes inside the hadron, where $m$ is a typical hadronic scale.
Within the \emph{pure} parton model, it is assumed that the hard
scattering is a lowest-order graph only (with exchange of a virtual
photon), without higher-order strong-interaction corrections.  Then
the cross section is schematically $\diff{\sigma} = \mbox{hard sc.}  \otimes
\mbox{pdf}$.

The space-time structure is illustrated in Fig.\ \ref{fig:DIS.sp.time}
in the Breit frame, i.e., where the exchanged virtual photon and the
proton are in the $(t,z)$ plane. The processes that lead to the
hadrons in the final state are dominantly roughly on a space-like
hyperbola at an invariant distance of order $1/m$ from the hard
scattering.  The hadronization region is indicated by the series of
short diagonal green lines.

Hadronization happens much too late to affect the fully inclusive
cross section; it just re-arranges the final state from a struck quark
and the proton remnant to something else.  Thus the parton-model cross
section is correctly given by the product of hard scattering and
parton density, without any allowance for the hadronization
processes. Notice how the validity of this argument relies both on the
use of unitarity for the final-state interactions and on a
coordinate-space picture that arises from the highly relativistic
kinematics.

\subsection{Fragmentation in $e^++e^- \to h_1 + h_2 + X$ at high $Q$}

\begin{figure}
\setlength\tabcolsep{0.025\textwidth}
\centering
\begin{tabular}{cc}
      \VC{\includegraphics[scale=0.53]{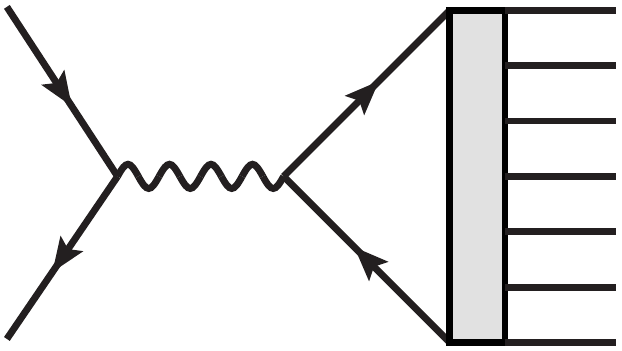}}
   &
      \VC{\includegraphics[scale=0.9]{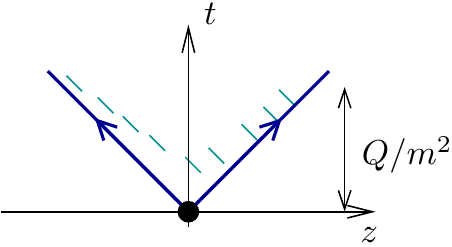}}
\\
   \begin{minipage}[t]{0.45\textwidth}
      \centering
      \caption{Amplitude for $e^+e^-\to\mbox{hadrons}$ to lowest order
      in hard scattering, with an indication of hadronization.}
      \label{fig:epem.basic}
   \end{minipage}
&
   \begin{minipage}[t]{0.45\textwidth}
      \caption{Space-time view corresponding to
        Fig.\ \protect\ref{fig:epem.basic}.}
      \label{fig:epem.string}
   \end{minipage}
\end{tabular}
\end{figure}

To examine the hadronization process in more detail, we consider
$e^+e^-\to\mbox{hadrons}$ at center-of-mass energy $Q$.  We use a
parton-model-like approximation, illustrated in
Fig.\ \ref{fig:epem.basic}, with its lowest order hard-scattering.  A
quark-antiquark pair is created over a short distance scale, of order
$1/Q$, which is followed by their propagation and hadronization, with
the space-time structure indicated in Fig.\ \ref{fig:epem.string}.
The two-particle-inclusive (2PI) cross section with hadrons in
opposite hemispheres is sensitive to the hadronization process, and
will bring in the relation to ffs.

A well-motivated model for hadronization is the string model
\cite{Artru:1974hr, Andersson:1983ia, Artru:1979ye,
  Andersson:1998tv}. Between the outgoing quark and antiquark is
generated a gluonic color flux tube.  Quark-antiquark pairs are
generated in the flux tube, and these rearrange themselves into
color-singlet hadrons, mostly pions; these fill in the rapidity gap
between the quark and antiquark with an approximately uniform
distribution in rapidity.  The diagram in Fig.\ \ref{fig:epem.basic}
is rather misleading, since it ignores the fact that the flux tube is
created by gluon emission quite early compared with the $Q/m^2$
time-scale for the generation of the fastest hadrons.  The central
hadrons are generated relatively early, on a time-scale $1/m$.  The
space-time structure, Fig.\ \ref{fig:epem.string}, follows from
\cite{Artru:1974hr, Andersson:1983ia, Artru:1979ye, Andersson:1998tv}.

The fastest oppositely moving hadrons are generated at space-like
separation. This and string-like fragmentation implies independent
fragmentation on the two sides, and so for hadrons in opposite
hemispheres, the approximations used so far can reasonably be expected
to give a factorized 2PI cross section with ffs, of the schematic form
\begin{equation}
    \frac{ \diff{\sigma} }{ \diff{z_1} \diff{z_2} }
    = \mbox{hard sc.} \otimes \mbox{ff}_1(z_1) 
                      \otimes \mbox{ff}_2(z_2)
      ~+~ \mbox{power suppressed}.
\label{eq:2PI.coll}
\end{equation}
However, because the rapidity gap is filled in, there are
Feynman-graph-like structures connecting the two opposite jets.

\section{Standard proofs of factorization}

For a systematic treatment of factorization and its derivation, see
\cite{Collins:2011qcdbook}, where also references to previous
literature can be found.

\subsection{Overview}

Given some suitable process with a large scale $Q$, the overall aims
are to extract the leading-power large-$Q$ asymptote of each
individual graph for process, and then to obtain a factorized form
after a sum over graphs.  (Sometimes a non-leading power is also
treated.)  This is carried out, in typical proofs, along the following
lines:
\begin{enumerate}
\item The regions that give the leading large-$Q$ behavior are
  determined. 
\item For each region, appropriate valid approximations are
  applied. The aims are that
  \begin{itemize}
  \item They should be suitable for obtaining a factorized cross
    section.
  \item They should allow the use of the Ward identities, as noted
    below, to combine graphs.
  \end{itemize}
\item For each term for each region, subtractions are applied to avoid
  double counting, etc.
\item Ward identities are applied to extract gluons connecting
  subgraphs for different classes of momenta (i.e., soft-to-collinear,
  collinear-to-hard).
\item Any relevant final-state unitarity cancellations are proved.
\item The result is a factorized form after a sum over graphs.
\end{enumerate}
 Evolution equations (DGLAP, etc) are derived similarly.

\begin{figure}
\setlength\tabcolsep{0.01\textwidth}
\centering
   \begin{tabular}{cc}
      \VC{\includegraphics[scale=0.5]{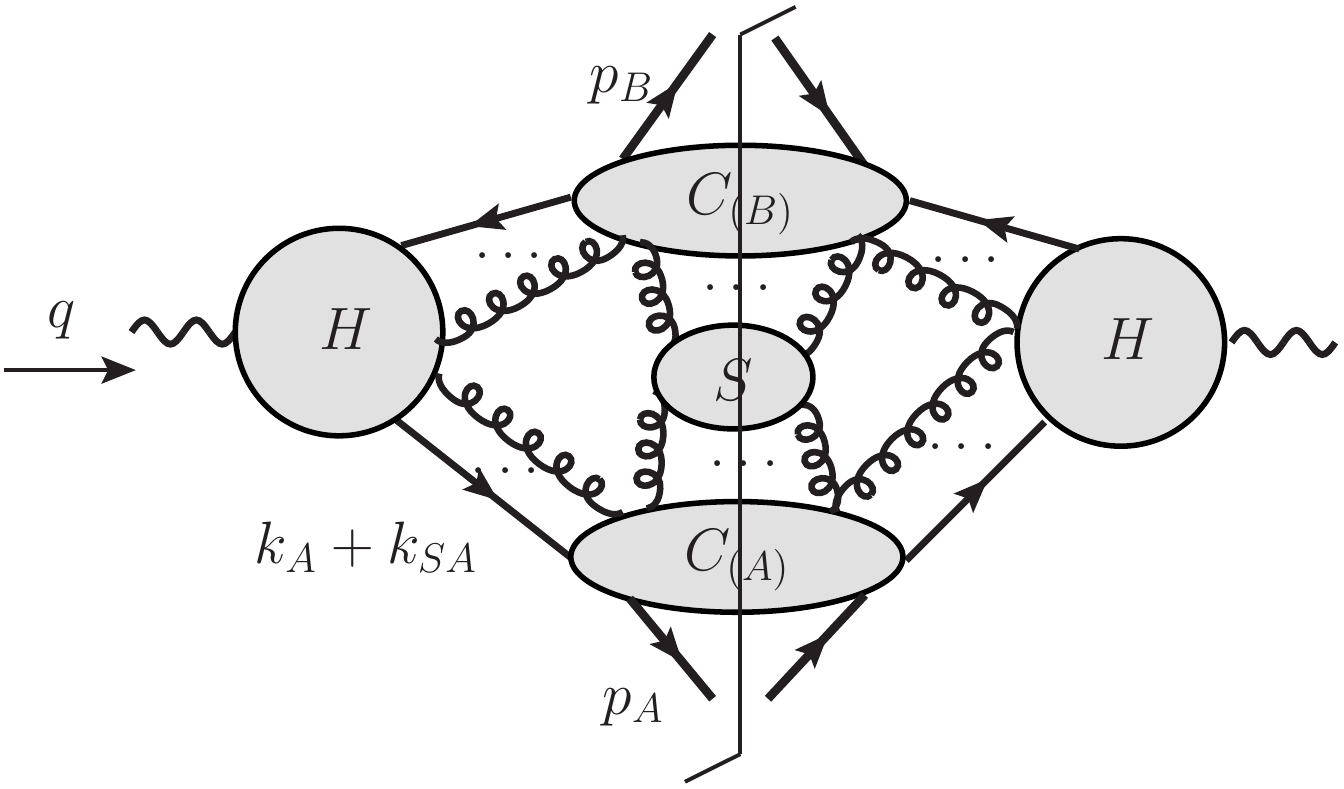}}
      &
      \VC{\includegraphics[scale=0.5]{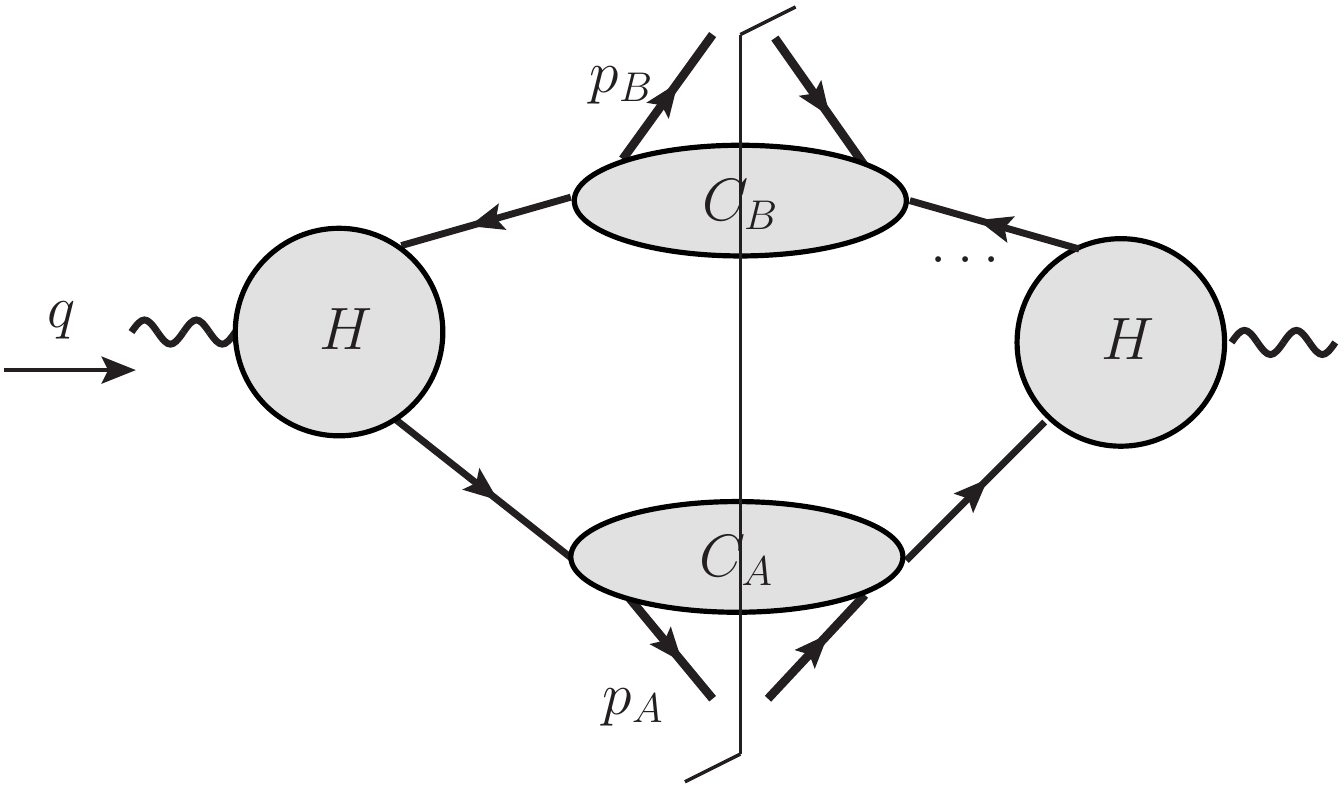}}
   \\
      (a) & (b)
   \end{tabular}
   \caption{Leading regions for $e^++e^- \to h_1 + h_2 + X$: (a) in
     full QCD; (b) in a parton-model-like situation.  These are for
     the case that the hadrons are close to back-to-back. The labels
     for the momentum classes of the subgraphs are: $H$ for hard, $S$
     for soft, and $C_{(A)}$ and $C_{(B)}$ for collinear momenta.}
\label{fig:epem.regions}
\end{figure}

The leading regions for the 2PI cross section in
$e^+e^-$-an\-nihi\-lation are shown in Fig.\ \ref{fig:epem.regions}.
To avoid the complications associated with emission of extra partons
at wide angle, I restrict attention to those regions that are leading
when the measured hadrons are almost back-to-back; this is a situation
appropriate for TMD factorization, with TMD ffs.

Figure \ref{fig:epem.regions}(a) shows the general QCD case.  Each
subgraph is associated with a momentum class: collinear to one or
other detected hadron, soft, or hard.  The simplest characterization
of the momentum classes is by the rough sizes, or scaling with $Q$, of
momentum components in light-front [$(+,-,T)$] coordinates.  For
example, the canonical scalings of collinear momenta and soft momenta
are:
\begin{equation}
  \mbox{Coll.\ A}: \left( Q, m^2/Q, m \right),
\qquad
  \mbox{S}: (m, m, m) ~\mbox{or}~ \left (m^2/Q, m^2/Q, m^2/Q \right).
\label{eq:scaling}
\end{equation}
The alternative soft scalings are called soft and ultrasoft, for two
different kinds of physics.  Almost completely, the proofs in
\cite{Collins:2011qcdbook} are arranged to work uniformly for soft and
ultrasoft. Although the scalings in (\ref{eq:scaling}) are very useful
to gain intuition about the regions, it is important that \emph{all
  intermediate momenta between the canonical ones also contribute at
  leading power}, and the proofs and the subtraction procedures must
be arranged \cite{Collins:2011qcdbook} to deal with this correctly.

Considerable complications in the proof arise because of the
arbitrarily many gluons connecting the soft subgraph to the collinear
subgraphs, and the collinear subgraphs to the hard subgraph.  Only
particular gluon polarizations contribute to leading power, and then
certain kinds of Ward identities are used to obtain the factorized
result, with gauge-invariant definitions of the factors.

If one had a theory without vector gluons, the leading regions would
be of the much simpler form of Fig.\ \ref{fig:epem.regions}(b).  Here
one already has a factorized form: a hard factor and two collinear
factors.  With appropriate approximations valid to leading power, the
collinear factors become the TMD ffs in the natural TMD generalization
of Eq.\ (\ref{eq:2PI.coll}) for the back-to-back region.

Several issues can be identified as to whether a proof using
order-by-order asymptotics in perturbation theory is sufficient:
\begin{itemize}
   \item Perturbation theory is not literally convergent.
   \item Non-perturbative effects exist in QCD.
   \item Even if one ignores the non-convergence issue, the large-$Q$
     asymptote and the sum of all orders of perturbation theory might
     not commute.
\end{itemize}
The last issue is rather important.  As we will see, as $Q$ is
increased, there is a set of graphs that are suggestive of the physics
of string fragmentation.  These are of order proportional to the
available rapidity range, i.e., to $\ln(Q^2/m^2)$, and they result in
a filled-in rapidity gap.  The asymptotics of each of these graphs is
only reached when $Q$ is much larger than the value where the rapidity
gap is filled.

\subsection{Non-perturbative-compatible structures in factorization
  proof}
\label{sec:non.pert.struct}

For a general graph the leading-power analysis, following the
Libby-Sterman \cite{Libby:1978bx, Libby:1978qf} argument, gives the
structure in Fig.\ \ref{fig:epem.regions}(a).  It is natural to think
of the blobs (or blocks) there, which are general subgraphs with
particular momentum classes, as being (subtracted) matrix elements in
full QCD, beyond perturbation theory.  Furthermore, a space-time
analysis shows that they essentially match a kind of (extended)
parton-model view, in coordinate space. Approximations can be applied
block-by-block rather than just
individual-subgraph-by-individual-subgraph.  In a case when no extra
gluonic connections exist, we get Fig.\ \ref{fig:epem.regions}(b),
where the approximated collinear blobs directly correspond to ffs in
the (TMD version of) factorization.

For these reasons, it would appear natural to expect that
factorization has a broader generality than just order-by-order in
perturbation theory.

The simplest parton-model-like regions, i.e.,
Fig.\ \ref{fig:epem.regions}(b), lead to two separated jets, with a
rapidity gap, of size approximately $\ln(Q^2/m^2)$, between them.
Moreover, in that situation, the jets have quark quantum numbers.
Thus, it is wrong to take Fig.\ \ref{fig:epem.regions}(b) at all
literally in QCD, since we know in reality large rapidity gaps are
normally filled in.

Finally, we note that in the Drell-Yan process, there is a highly
non-trivial cancellation of spectator-spectator interactions.  The
earliest proofs of cancellation, which essentially pre-date QCD, were
fully non-perturbative \cite{DeTar:1974vx, Cardy:1974vq}; but these
proofs used strict parton-model assumptions that are false in
QCD. Collins, Soper and Sterman \cite{Collins:1985ue, Collins:1988ig}
extended that work to full QCD, with an updated version in
\cite{Collins:2011qcdbook}.

\section{What's missing?}

\subsection{Perturbative proof v.\ filled-in rapidity gap}

Suppose Fig.\ \ref{fig:epem.regions}(b), were sufficient.  Then we
could interpret the blobs as full non-perturbative quantities.  The
natural leading-power approximations lead to factorization, complete
with the natural operator definitions of the ffs. Even before the
advent of QCD and the modern understanding of factorization,
essentially these arguments were applied to relate the parton model
for DIS to a full field theory of the strong interaction
\cite{Landshoff:1971xb} (but with assumptions that are not valid in
QCD).

We need something more general in QCD.  Within the order-by-order
analysis, we have Fig.\ \ref{fig:epem.regions}(a).  The approximations
that lead to factorization require large rapidity differences between
the subgraphs.  For example, consider a soft gluon of momentum $k$
attaching to the collinear-to-A subgraph.  With the canonical powers
(\ref{eq:scaling}), the appropriate approximation for the product of
collinear and soft subgraphs is made by
\begin{equation}
  A(k,\dots)^\mu S(k,\dots)_\mu
  \simeq
  A(\hat{k},\dots)^+ S(k,\dots)^-
  =
  A(\hat{k},\dots)\cdot \hat{k} \frac{1}{k^-} S(k,\dots)^- ,
\label{eq:soft.approx}
\end{equation}
\begin{wrapfigure}{l}{0.43\textwidth}
   \centering
   \VC{\includegraphics[scale=0.38]{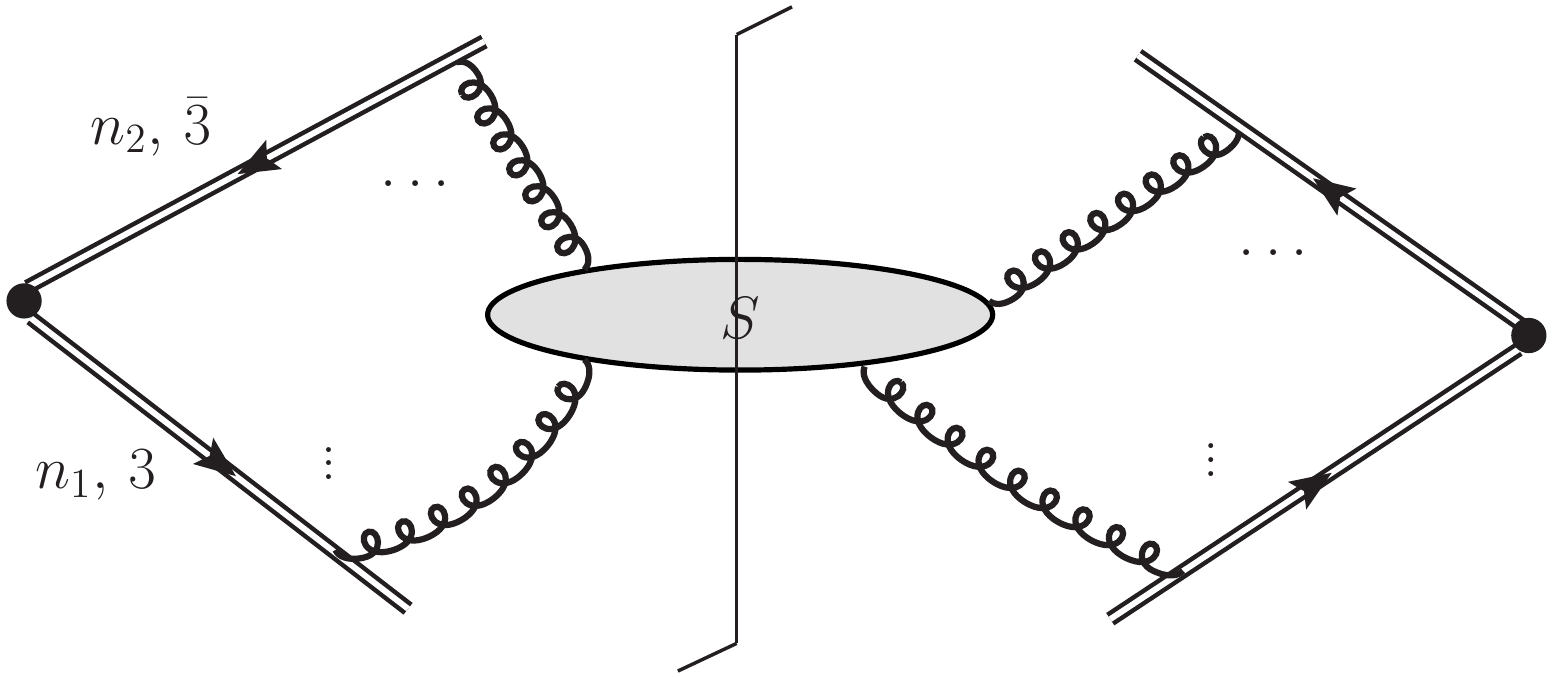}}
\caption{Generic soft factor.}
\label{fig:soft}
\end{wrapfigure}
where $\hat{k}= (0,k^-,\T{0})$ in light-front $(+,-,T)$ coordinates,
and we define collinear-to-A momenta to have large $+$ components.
The approximated graph has a factor of approximated gluon momentum
contracted with the collinear subgraph with the same approximated
momentum, i.e., $A(\hat{k},\dots)\cdot \hat{k}$.  There exists a Ward
identity to convert this soft-gluon attachment, and all the others in
Fig.\ \ref{fig:epem.regions}(a), to the form of a factorized soft
piece that is a matrix element of Wilson lines, illustrated in
Fig.\ \ref{fig:soft}.  Essentially the same argument applies to the
gluons connecting collinear to hard subgraphs, and results in the
Wilson lines in the operator definition of the ffs.

A full treatment \cite{Collins:2011qcdbook} needs to take account of
subtractions and various kind of divergence; but the core of the
argument consists of Eq.\ (\ref{eq:soft.approx}).  Furthermore, in TMD
factorization, it turns out to be useful to absorb
\cite{Collins:2011qcdbook, GarciaEchevarria:2011rb} square roots of
the soft factor into the definitions of the collinear factors, i.e.,
the ffs, or more generally the pdfs and the ffs.  Also, in collinear
factorization a unitarity cancellation converts the soft factors into
unity.

However, in reality the rapidity gaps are filled in, with about two
particles per unit rapidity.  So the assumption used in
(\ref{eq:soft.approx}), i.e., of large rapidity differences, is not
always true in real QCD; the assumption is appropriate only for the
large-$Q$ asymptotics for an (\emph{arbitrary}) individual graph.

\subsection{Soft-to-collinear approximation fails in
  string-compatible graphs}

We can even see the failure of the proof in a perturbative model that
mimics the physics of string hadronization, as used in the PYTHIA
event generator \cite{Sjostrand:2006za}, or of a similar cluster
hadronization used in HERWIG \cite{Corcella:2000bw}. (See
\cite{Webber:1999ui} for a review of hadronization and fragmentation.)

\begin{wrapfigure}{l}{0.40\textwidth}
   \centering
   \includegraphics[scale=0.75]{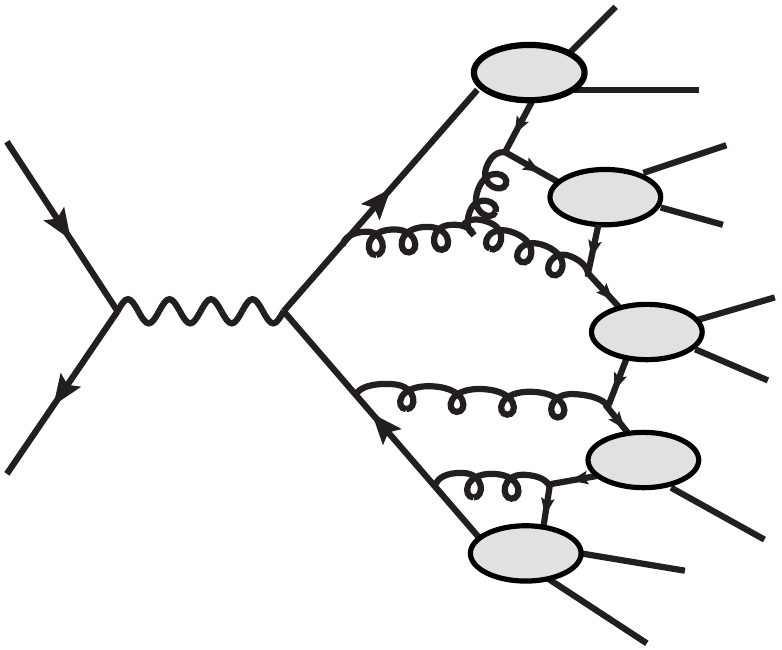}
\caption{Cluster hadronization.  The number of blobs for the final
  state interactions is proportional to $\ln(Q^2/m^2)$.}
\label{fig:cluster}
\end{wrapfigure}

The model is illustrated in Fig.\ \ref{fig:cluster}.  It corresponds
to the following semi-intuitive description: We have a quark and
antiquark that are generated at a small distance and that propagate
outward in opposite directions with energy approximately $Q/2$, and
with low virtuality.  At early stages, gluons are emitted, and then
quark-antiquark pairs are generated, which organize themselves into
color-singlet hadrons.  This is all qualitatively like the string
model.  Since virtualities decrease as we go away from the hard
scattering, the relevant effective coupling increases (and ultimately
we get to a non-perturbative region).  Note that important
contributions to the momenta of the central hadrons are from the soft
gluons that are emitted early.

Let $\Delta y$ be total available rapidity range, about
$\ln(Q^2/m^2)$, and let $\delta y$ be typical cluster separation,
i.e., $\Delta y/\#\mbox{clusters}$.  Experimentally $\delta y$ is
typically small, a unit of rapidity or so.

Now suppose we were able apply soft-to-collinear approximations.
Errors would now be a power of $e^{-\delta y}$, not $m/Q$, while the order
of the relevant graphs increases with $\ln Q$.  Although the middle
blobs in Fig.\ \ref{fig:cluster} have soft momenta with momenta of
order $(m,m,m)$, their structure fails to match Fig.\
\ref{fig:epem.regions}(a) with its purely gluonic connections, and we
do not have the rapidity differences needed in (\ref{eq:soft.approx}).
In contrast, the soft-to-collinear approximation does apply to the
gluons that are emitted early.

We thus see that the proof of factorization, as presented in the
literature, does not work in graphical structures that are clearly
important for getting the correct final state.  

The above considerations do not themselves entail that factorization
fails or even that the definitions of the pdfs and ffs are wrong, but
only that there are some critical issues that are not fully
understood.  As already pointed out, the space-like separation between
the dominant hadronization processes in different jets strongly
suggests that some kind of factorization is valid, with something like
ffs.  But the coordinate-space description does not itself allow to
obtain operator definitions of the ffs in a simple way.  Furthermore,
such coordinate-space arguments are quite foreign to the
momentum-space methods used in factorization proofs.

A more indirect approach might be fruitful.  The coordinate-space
factorization argument could be applied to the operator matrix
elements in the definition of ffs, since these involve color flows in
very different directions.  From this approach one may well recover
the usual factorization results, but now demonstrated to be valid in
full QCD.  But completing such a program is for the future.

\section{Discussion}

The overall issue addressed in this paper is: How well do we know that
factorization is valid?  Then, given that we do find that the
derivations need extensions, at least concerning hadronization, a
question for the future is to ask what new phenomena and properties
could be found from a better understanding.

We saw that string-type hadronization \cite{Webber:1999ui} does not
match perturbative derivations of factorization, even in a
perturbative model. This specific problem applies everywhere that
cross sections are differential in final-state particles.  This is
only one part of the overall issue of asking how to establish
factorization theoretically within full QCD as opposed to literally
order-by-order in perturbation theory.  Even so, as I pointed out in
Sec.\ \ref{sec:non.pert.struct}, there are many parts of the current
proofs that do transcend a purely perturbative treatment.

The arguments in this paper do not themselves imply that factorization
is falsified.  The opposite ends of a Lund string are at space-like
separation, so we can still expect independent hadronization (except
for the effects of QM entanglement).  There is tension or even a
clash, however, between the coordinate-space picture that gives this
expectation, and the momentum-space formulation of most work on
factorization.

There are a number of directions for future work that seem useful:
\begin{itemize}
\item Do the issues I raised actually impact factorization formulas
  with fragmentation functions?  This applies to both collinear and
  TMD factorization.
\item We need to find an improved formulation, perhaps involving a
  proper systematic interface between standard pQCD constructs and
  string-like constructs.
\item What other similar issues affect derivations of factorization,
  including those with pdfs?
\item We need improved and quantitative methods for analyzing graphs
  and amplitudes in coordinate space, and for relating them to work in
  momentum space.
\item What other implications are there?
\item What other problems are there?
\item How can one understand better the relation with the methods of
  soft-collinear effective theory?
\end{itemize}
A better formulation should assist in better matching the algorithms
used in Monte-Carlo event generators to the TMD factorization
formalism.



\providecommand{\href}[2]{#2}
\begingroup\raggedright

\endgroup

\end{document}